\documentclass[smallextended]{svjour3}       
\usepackage{graphicx}
\usepackage{bbm}
\usepackage{mathptmx}      
\usepackage{latexsym}
\journalname{Quantum Information Processing}

\begin{document}

\title{A simpler algorithm to mark the unknown eigenstates}
\author{Avatar Tulsi}
\institute{Avatar Tulsi \at
              Department of Physics, IIT Bombay, Mumbai - 400076, India \\
              Tel.: +91-22-2576-7596\\
              \email{tulsi9@gmail.com}}

\date{Received: date / Accepted: date}

\maketitle

\begin{abstract}

For an unknown eigenstate $|\psi\rangle$ of a unitary operator $U$, suppose we have an estimate of the corresponding eigenvalue which is separated from all other eigenvalues by a minimum gap of magnitude $\Delta$. In the eigenstate-marking problem (EMP), the goal is to implement a selective phase transformation of the $|\psi\rangle$ state (known as \emph{marking} the $|\psi\rangle$ state in the language of the quantum search algorithms). The EMP finds important applications in the construction of several quantum algorithms. The best known algorithm for the EMP combines the ideas of the phase estimation algorithm and the majority-voting. It uses $\Theta(\frac{1}{\Delta}\ln \frac{1}{\epsilon})$ applications of $U$ where $\epsilon$ is the tolerable error. It needs $\Theta\left(\ln \frac{1}{\Delta}\right)$ ancilla qubits for the phase estimation and another $\Theta\left(\ln \frac{1}{\epsilon}\right)$ ancilla qubits for the majority-voting. 

	In this paper, we show that the majority-voting is not a crucial requirement for the EMP and the same purpose can also be achieved using the fixed-point quantum search algorithm which does not need any ancilla qubits. In the case of majority-voting, these ancilla qubits were needed to do controlled transformations which are harder to implement physically. Using fixed-point quantum search, we get rid off these $\Theta\left(\ln \frac{1}{\epsilon}\right)$ ancilla qubits and same number of controlled transformations. Thus we get a much simpler algorithm for marking the unknown eigenstates. However, the required number of applications of $U$ increases by the factor of $\Theta \left(\ln \frac{1}{\epsilon}\right)$. This tradeoff can be beneficial in typical situations where spatial resources are more constrained or where the controlled transformations are very expensive.
\end{abstract}

\section{Introduction}
\label{introduction}

	Suppose we have a unitary operator $U$ and an estimate of one of its eigenvalues, say $e^{\imath \psi}$. Let the corresponding eigenstate be $|\psi\rangle$ which is not known to us. We consider the problem of implementing a selective phase transformation $I_{\psi}^{\phi}$ which applies a phase factor of $e^{\imath \phi}$ to the $|\psi\rangle$ state but leave all other eigenstates $|\psi_{\perp}\rangle$ of $U$ (orthogonal to $|\psi\rangle$) unchanged. Mathematically, $I_{\psi}^{\phi}|\psi\rangle$ is $e^{\imath \phi}|\psi\rangle$ whereas $I_{\psi}^{\phi}|\psi_{\perp}\rangle$ is $|\psi_{\perp}\rangle$. We call this problem the \emph{eigenstate-marking} problem (hereafter referred to as EMP) as in the language of the quantum search algorithms~\cite{grover1,grover2}, applying the selective phase transformation of an unknown quantum state is also known as \emph{marking} that state.	

	The EMP finds applications in the construction of several important quantum algorithms. For example, in the quantum principal component analysis having applications to the pattern and face recognition~\cite{qpca1}, the EMP is used to mark the eigenstates corresponding to the largest (principal) eigenvalues of an operator. In the Eigenpath Traversal Problem, the EMP is used for the selective phase inversion of intermediate eigenstates without which the efficient algorithms~\cite{BKSpaper,PEAapprox} do not work. Note that the quantum adiabatic evolution is a special case of the Eigenpath traversal problem. Thus EMP helps in getting a faster simulation of quantum adiabatic evolution as explicity discussed in ~\cite{digital}. The EMP is also needed for optimal quantum measurements of expectation values of observables as shown in ~\cite{highconfidence}. The EMP also finds applications in the quantum post-processing to speed up general quantum search algorithms~\cite{general} as shown in ~\cite{postprocessing}.

	All algorithms for the EMP is based on the following basic idea. We attach an ancilla work-space with the Hilbert space $\mathcal{H}_{w}$ to the main space $\mathcal{H}_{m}$ (the Hilbert space of the main quantum system) on which we wish to implement $I_{\psi}^{\phi}$. We work in the joint space $\mathcal{H}_{j} = \mathcal{H}_{m} \otimes \mathcal{H}_{w}$. Upt to an error of $O(\epsilon)$, implementation of $I_{\psi}^{\phi}$ is equivalent to implementing an operator $C_{\epsilon}$ which satisfies the following:
\begin{eqnarray}
C_{\epsilon}\left(|\psi\rangle |\sigma\rangle \right) & = & |\psi\rangle \left(|Z\rangle + |\epsilon\rangle\right), \nonumber \\
C_{\epsilon}\left(|\psi_{\perp}\rangle |\sigma\rangle \right) & = & |\psi_{\perp}\rangle \left(|Z^{\perp}\rangle + |\epsilon\rangle\right).  \label{Cdefine} 
\end{eqnarray}
Here $|\sigma\rangle$ is any standard known state of $\mathcal{H}_{w}$. The $|Z\rangle$ and $|Z^{\perp}\rangle$ states are completely within the two mutually complementary and easily distinguishable subspaces of $\mathcal{H}_{w}$ which we denote by $\mathcal{Z}$ and $\mathcal{Z}^{\perp}$ respectively. The $|\epsilon\rangle$ denotes any arbitrary state of length smaller than $\epsilon$ where $\epsilon \ll 1$ is the tolerable implementation error. For a perfect implementation of $I_{\psi}^{\phi}$, $\epsilon$ is zero. But a perfect implementation is generally not possible and only approximate implementation can be achieved.  

	A naive approach to implement $C_{\epsilon}$ is by implementing $P_{\epsilon}$, the operator corresponding to the phase estimation algorithm (hereafter referred to as PEA)~\cite{phase}. The operator $P_{\epsilon}$ uses $\mu$ ancilla qubits for the work-space and $\Theta(2^{\mu})$ applications of the operator $U$. The error term $\epsilon$ is $O(1/\sqrt{2^{\mu}\Delta})$ where $\Delta$ is the minimum value of $|\psi_{\perp} -\psi|$ among all $\psi_{\perp} \neq \psi$ (we consider only the typical cases when $\Delta \ll 1$). For any operator $X$, let $\mathcal{N}_{U}(X)$ and $\mathcal{N}_{A}(X)$ denote the $U$-complexity (the required number of applications of $U$ or $U^{\dagger}$) and the ancilla-complexity (the required number of ancilla qubits) respectively. As $\epsilon$ is $O(1/\sqrt{2^{\mu}\Delta})$ for $P_{\epsilon}$, it is easy to check that 
\begin{eqnarray}
\mathcal{N}_{U}(P_{\epsilon}) &=& 2^{\mu} =  \Theta\left(\frac{1}{\Delta \epsilon^{2}}\right), \nonumber \\
\mathcal{N}_{A}(P_{\epsilon}) &=& \mu = \Theta\left(\log_{2}\frac{1}{\Delta \epsilon^{2}}\right)  = \Theta\left(\log_{2}\frac{1}{\Delta} + 2\log_{2}\frac{1}{\epsilon}\right). \label{NUAP}
\end{eqnarray}
The desired error $\epsilon$ in each implementation of $I_{\psi}^{\phi}$ depends upon $Q$, the total number of implementations required by any procedure. We must have $\epsilon \ll 1/Q$ for the success of procedure as the errors add up linearly in quantum mechanics. In many applications, the operator $P_{\epsilon}$ becomes highly inefficient as $\epsilon$ is too small making $\mathcal{N}_{U}(P_{\epsilon})$ large enough to make the entire procedure inefficient. This has been discussed earlier in several papers (for example, see ~\cite{highconfidence,postprocessing}).

	To overcome the inefficiency of $P_{\epsilon}$, a high-confidence version of the PEA was presented in ~\cite{highconfidence}. Here the operator $C_{\epsilon}$ is chosen to be $H_{\epsilon} = P_{1/32}^{\otimes \nu}$ which is a parallel application of $P_{1/32}$ on all $\nu$ quantum registers. Here $P_{1/32}$ is $P_{\epsilon}$ with a constant error parameter $\epsilon = 2^{-5}$. To implement $H_{\epsilon}$, we use multiple ($\nu$) quantum registers where each register consists of $\Theta(\ln \frac{1}{\Delta})$ ancilla qubits required by $P_{1/32}$ as per Eq. (\ref{NUAP}). We use the majority-voting after application of $H_{\epsilon}$ as we check that whether majority of the $\nu$ registers are in a particular known state (subspace) or not. The number $\nu$ is decided by the tolerable error $\epsilon$ which decreases exponentially with $\nu$ as $\epsilon = O(e^{-\nu/4})$. So $\nu$ is $\Theta\left(\ln \frac{1}{\epsilon}\right)$ and the complexities of the operator $H_{\epsilon}$ is given by
\begin{eqnarray}
\mathcal{N}_{U}(H_{\epsilon})& = & \Theta\left(\frac{1}{\Delta}\ln\frac{1}{\epsilon}\right), \nonumber \\
\mathcal{N}_{A}(H_{\epsilon}) &=& \Theta\left(\ln\frac{1}{\epsilon}\ln \frac{1}{\Delta}\right). \label{NUAH}
\end{eqnarray} 
Comparing this to Eq. (\ref{NUAP}), we find that $\mathcal{N}_{U}(H_{\epsilon}) \ll \mathcal{N}_{U}(P_{\epsilon})$ for $\epsilon \ll 1$. Thus we significantly save the required number of applications of $U$. But this saving comes at the expense of huge number of ancilla qubits as $\mathcal{N}_{A}(H_{\epsilon})$ is the product of $\Theta\left(\ln\frac{1}{\epsilon}\right)$ and $\Theta\left(\ln \frac{1}{\Delta}\right)$ whereas $\mathcal{N}_{A}(P_{\epsilon})$ is only the sum of these two terms. Hence the required number of ancilla qubits becomes too large for small values of $\epsilon$.

	Recently, a modified high-confidence version of the PEA was presented in ~\cite{postprocessing} with a much better ancilla-complexity. This modification is based on the idea that for majority-voting, we are not interested in the exact quantum states of all $\nu$ registers but only their components in two mutually complementary and known subspaces $\mathcal{Z}$ and $\mathcal{Z}^{\perp}$. These two subspaces can be represented by the two basis states $|0\rangle$ and $|1\rangle$ of an ancilla qubit. The modified algorithm then simulates the quantum state of each register by this ancilla qubit. Thus the quantum state of all $\nu$ registers is simulated using $\nu$ ancilla qubits. For this simulation, we require only one register. Hence we remove the necessity of multiple quantum registers by simulating their states using ancilla qubits with one ancilla qubit per register. Let $M_{\epsilon}$ denote the operator corresponding to the modified algorithm. On the one hand, $\mathcal{N}_{U}\left(M_{\epsilon}\right) \approx \mathcal{N}_{U}\left(H_{\epsilon}\right)$ as far as the $U$-complexity is concerned but on the other hand, $\mathcal{N}_{A}\left(M_{\epsilon}\right) \approx \mathcal{N}_{A}\left(P_{\epsilon}\right)$ as far as the ancilla-complexity is concerned. Thus $M_{\epsilon}$ retains the advantageous features of both $P_{\epsilon}$ and $H_{\epsilon}$ while discarding their disadvantageous features.        

	In this paper, we show that the combination of the phase estimation algorithm and the majority-voting is not the only way to get an efficient algorithm for the EMP. A better and simpler algorithm is available if we use the phase estimation algorithm along with the fixed-point quantum search algorithm~\cite{fixed1,fixed2} which ensures the monotonic convergence of a particular quantum state towards another quantum state. Let $F_{\epsilon}$ denote the operator corresponding to fixed-point quantum search based algorithm. We show that its complexities are given by
\begin{eqnarray}
\mathcal{N}_{U}(F_{\epsilon}) &=& \Theta\left(\frac{1}{\Delta}\ln^{2}\frac{1}{\epsilon}\right), \nonumber \\
\mathcal{N}_{A}(F_{\epsilon}) &=& \Theta\left(\ln \frac{1}{\Delta}\right). 
\end{eqnarray}  
Thus the required number of ancilla qubits is completely independent of the tolerable error. The ancilla qubits are basically needed to do controlled transformations which are harder to implement physically. Using the fixed-point quantum search, we get rid off the ancilla qubits used for majority-voting and thus we get rid off significant number of controlled transformations. Hence we get a much simpler algorithm. This comes at the cost of the increase of the number of applications of $U$ by a factor of $\ln\frac{1}{\epsilon}$. This tradeoff is beneficial in typical situations when the spatial resources are constrained or when the controlled transformations are very expensive. 

	The paper is organized as follows. In the next section, we briefly review the majority-voting based approaches to the EMP. In the Section III, we present the simpler algorithm based on the fixed-point quantum search. We discuss and conclude in Section IV.

\section{Majority-Voting based algorithms}
\label{majorityvoting}

	We briefly review the earlier algorithms which uses the majority-voting to enhance the efficiency of the phase estimation algorithm for the EMP. For more details, we refer the readers to the Section III of ~\cite{postprocessing} (the notation used there is slightly different from the notation used here). A detailed algorithm has been presented there for implementation of $I_{\lambda_{\pm}}$, the selective phase inversion of the two eigenstates $|\lambda_{\pm}\rangle$ of an operator $\mathcal{S}$ corresponding to the eigenvalues $e^{\imath \lambda_{\pm}}$. The only assumption is that the two eigenphases $\lambda_{\pm}$ satisfy $|\lambda_{\pm}| \ll \theta_{\rm min}$ whereas all eigenphases $\lambda_{\perp}$ other than $\lambda_{\pm}$ satisfy $|\lambda_{\perp}| > \theta_{\rm min}$. If $\lambda_{+}$ and $\lambda_{-}$ have a common value $\lambda$ then the algorithm of ~\cite{postprocessing} can be used to approximate $I_{\lambda}$, the selective phase inversion of an eigenstate $|\lambda\rangle$ of $\mathcal{S}$ provided the assumption 
\begin{equation}
|\lambda| << \theta_{\rm min},\ \ |\lambda_{\perp}| > \theta_{\rm min} \label{assumption1}
\end{equation}
is satisfied.

	In the EMP, we want to implement $I_{\psi}^{\phi}$, the selective phase rotation of an eigenstate $|\psi\rangle$ of a unitary operator $U$ corresponding to the eigenvalue $e^{\imath \psi}$. The magnitude of the difference between the eigenphase $\psi$ and other eigenphases $\psi_{\perp}$ of $U$ is at least $\Delta$, i.e. $|\psi_{\perp} - \psi| > \Delta$. Suppose we have an estimate of $\psi$ up to an accuracy of $b\Delta$ where $b \ll 1$ is a small constant. This means that we know a number $\psi'$ such that $|\psi' - \psi| < b\Delta$. Using this knowledge, we can easily construct a unitary operator $\mathcal{S} = e^{-\imath \psi'}U$. Then it is easy to check that the assumption (\ref{assumption1}) is satisfied if we choose the eigenstate $|\psi\rangle$ of $U$ to be the eigenstate $|\lambda\rangle$ of $\mathcal{S}$ and if we choose $\theta_{\rm min}$ to be $\Delta/2$. Thus we can use the framework presented in ~\cite{postprocessing} for the problem of eigenstate-marking.

	The implementation of $I_{\lambda}^{\phi}$ with an error of $O(\epsilon)$ is equivalent to the implementation of an operator $C_{\epsilon}$ satisfying Eq. (\ref{Cdefine}). As elaborated in the subsection III.A of ~\cite{postprocessing}, up to an error of $O(\epsilon)$, the desired operator $I_{\lambda}^{\phi}$ is given by  
\begin{equation}
C_{\epsilon}^{\dagger}\left(\mathbbm{1}_{m} \otimes I_{\mathcal{Z}}^{\phi}\right)C_{\epsilon} \label{Rdefine}
\end{equation}
where $\mathbbm{1}_{m}$ is the identity operator acting on the mainspace and $I_{\mathcal{Z}}^{\phi}$ is the selective phase rotation by an angle $\phi$ of the $\mathcal{Z}$-subspace of the work-space. Note that in ~\cite{postprocessing}, we have discussed only the special case when $\phi$ is $\pi$ but it is straightforward to extend the discussion to the general values of $\phi$. 

	A simple implementation of $C_{\epsilon}$ is to simulate it by the operator $P_{\epsilon}$ corresponding to the phase estimation algorithm. We choose a work-space $\mathcal{H}_{w}$ of $\mu$ ancilla qubits. The operator $P_{\epsilon}$ is a successive application of three operators, i.e. $P = P(3)P(2)P(1)$. The first operator $P(1)$ applies a Walsh-Hadamard transform on the workspace. The second operator $P(2)$ applies $\mathcal{S}^{z}$ on the mainspace if and only if the workspace is in the basis state $|z\rangle$. As the work-space of $\mu$ qubits is $2^{\mu}$-dimensional, $z$ ranges from $0$ to $2^{\mu}-1$. Hence $P(2)$ requires $2^{\mu}$ applications of $U$. The third operator $P(3)$ applies the inverse quantum fourier transform on the workspace. As shown in the subsections III.B and III.C of ~\cite{postprocessing}, by choosing the operator $C_{\epsilon}$ in Eq. (\ref{Cdefine}) to be $P_{\epsilon}$, we can choose $\epsilon$ to be $1/\sqrt{2^{\mu}\Delta}$. Thus the $U$-complexity and the ancilla-complexity of the operator $P_{\epsilon}$ is given by Eq. (\ref{NUAP}).

	In the high-confidence version of the PEA~\cite{highconfidence}, we use multiple ($\nu$) registers of ancilla qubits. We do parallel applications of the operator $P_{1/32}$ on all $\nu$ registers. Here $P_{1/32}$ is the operator $P_{\epsilon}$ with a constant error parameter $\epsilon = 2^{-5}$. Let the operator $H_{\epsilon} = P_{1/32}^{\otimes \nu}$ denote this parallel application. Then  
\begin{eqnarray}
H_{\epsilon}\left(|\psi\rangle |\sigma\rangle^{\otimes \nu} \right) & = & |\psi\rangle \left(|Z\rangle + |2^{-5}\rangle\right)^{\otimes \nu}, \nonumber \\
H_{\epsilon}\left(|\psi_{\perp}\rangle |\sigma\rangle^{\otimes \nu}\right) & = & |\psi_{\perp}\rangle \left(|Z^{\perp}\rangle + |2^{-5}\rangle\right)^{\otimes \nu}. 
\end{eqnarray}
Let $|>\rangle$ ($|<\rangle$) denote a normalized state in which more (less) than $\frac{\nu}{2}$ registers are in $|Z\rangle$ state. Equivalently, $|>\rangle$ ($|<\rangle$) denote a normalized state in which less (more) than $\frac{\nu}{2}$ registers are in $|Z^{\perp}\rangle$ state. 

	Suppose the main system is in $|\psi\rangle$ state. Then the probability of getting a single register in $|Z^{\perp}\rangle$ state is less than $2^{-10}$. Thus the expected number of registers in $|Z^{\perp}\rangle$ state is less than $2^{-10}\nu$. Due to Hoeffding's bound~\cite{hoeffding}, the probability of getting more than $\frac{\nu}{2}$ registers in $|Z^{\perp}\rangle$ state is at most $e^{-2\nu t^{2}}$ where $t$ is $\frac{1}{2}-O(2^{-10}) \approx \frac{1}{2}$. Thus the amplitude of the joint state of all $\nu$ registers in $|<\rangle$ state is less than $e^{-\nu/4}$. Similar considerations will show that if the main system is in $|\psi_{\perp}\rangle$ state then the amplitude of the joint state of all $\nu$ registers in $|>\rangle$ state is less than $e^{-\nu/4}$. Thus we get
\begin{eqnarray}
H_{\epsilon}\left(|\psi\rangle |\sigma\rangle^{\otimes \nu} \right) & = & |\psi\rangle \left(|>\rangle + |e^{-\nu/4}\rangle\right), \nonumber \\
H_{\epsilon}\left(|\psi_{\perp}\rangle |\sigma\rangle^{\otimes \nu}\right) & = & |\psi_{\perp}\rangle \left(|<\rangle + |e^{-\nu/4}\rangle\right). 
\end{eqnarray}
Comparing this with Eq. (\ref{Cdefine}) and by considering the basis $\{|>\rangle,|<\rangle\}$ in place of the basis $\{|Z\rangle,|Z^{\perp}\rangle\}$, we find that the error $\epsilon$ decreases exponentially with the number of registers as $e^{-\nu/4}$. Thus the required number of registers $\nu$ is $\Theta\left(\ln \frac{1}{\epsilon}\right)$.

	The number of ancilla qubits of each register is chosen as required by the operator $P_{1/32}$. Putting $\epsilon = 2^{-5}$ in eq. (\ref{NUAP}), we find that each register must have $\Theta\left(\ln\frac{1}{\Delta}\right)$ ancilla qubits and each implementation of $P_{1/32}$ requires $\Theta(1/\Delta)$ applications of $U$. As $\nu$ is $\Theta \left(\ln\frac{1}{\epsilon}\right)$, it is easy to check that the $U$-complexity and the ancilla-complexity of the operator $H_{\epsilon}$ are given by Eq. (\ref{NUAH}). Though the $U$-complexity is much better than $P_{\epsilon}$ but the ancilla complexity is very poor compared to $P_{\epsilon}$.

	The high-confidence version can be further modified and simplified to get a much better ancilla-complexity as done in the subsection III.D of ~\cite{postprocessing}. In this modification, we get rid off $\nu$ registers which is the main reason for the poor ancilla complexity of the operator $H_{\epsilon}$. We note that each register is independently used for the operator $P_{1/32}$ given by
\begin{eqnarray}
P_{1/32}\left(|\psi\rangle |\sigma\rangle \right) & = & |\psi\rangle \left(|Z\rangle + |2^{-5}\rangle\right), \nonumber \\
P_{1/32}\left(|\psi_{\perp}\rangle |\sigma\rangle \right) & = & |\psi_{\perp}\rangle \left(|Z^{\perp}\rangle + |2^{-5}\rangle\right).  \label{P132define} 
\end{eqnarray}   
We are only interested in the magnitudes of amplitudes of each register's state in two mutually orthogonal and easily distinguishable quantum states $|Z\rangle$ and $|Z^{\perp}\rangle$. 

	Suppose we have a single ancilla qubit whose basis states $\{|0\rangle,|1\rangle\}$ represent the basis $\{|Z\rangle,|Z^{\perp}\rangle\}$. The state of this qubit is either $|\beta_{\psi}^{\pm}\rangle$ or $|\beta_{\psi\perp}^{\pm}\rangle$ depending upon whether the state of the main space is $|\psi\rangle$ or $|\psi_{\perp}\rangle$ respectively. These qubit states are chosen to simulate the distribution of amplitudes of each register in $|Z\rangle$ and $|Z^{\perp}\rangle$ states. Thus  
\begin{eqnarray}
|\beta_{\psi}^{\pm}\rangle &=& \cos \beta_{\psi} |0\rangle \pm \imath \sin \beta_{\psi}|1\rangle,\ \ \sin \beta_{\psi} \leq 2^{-5}, \nonumber \\
|\beta_{\psi\perp}^{\pm}\rangle &=& \cos \beta_{\psi\perp} |0\rangle \pm \imath \sin \beta_{\psi\perp}|1\rangle,\ \ \cos \beta_{\psi\perp} \leq 2^{-5}.   
\label{betadefine} 
\end{eqnarray} 
The above single qubit state is a very important resource. As the qubit simulates the amplitude-distribution of a register, we need at least one register to get this qubit. We start with an ancilla qubit in a standard quantum state. Then we apply a controlled amplitude-amplification operator on the register. For this amplitude-amplification operator, we need $\Theta(1)$ applications of $P_{1/32}$. After this, we apply a Hadamard gate on the ancilla qubit. The ancilla qubit state becomes exactly the desired state (\ref{betadefine}) as shown in detail in the subsection III.D of ~\cite{postprocessing}. The same register can be used to get arbitrarily many copies of qubits in the desired state (\ref{betadefine}).
 	
	Suppose we have $\nu$ identical copies of such a qubit. Let $\mathcal{H}_{2^{\nu}}$ denote the joint Hilbert space of all $\nu$ qubits. Let $0^{+}$ denote a subspace of $\mathcal{H}_{2^{\nu}}$ in which at least $\nu/2$ qubits are in $|0\rangle$ state. Let $1^{+}$ denote the complementary subspace of $0^{+}$. Let $|0^{+}\rangle$ and $|1^{+}\rangle$ denote any normalized state completely within the subspaces $0^{+}$ and  $1^{+}$ respectively. By definition, the distribution of the amplitudes of $\nu$ qubits in the states $|0^{+}\rangle$ and $|1^{+}\rangle$ simulate the distribution of amplitudes of $\nu$ registers in the states $|>\rangle$ and $|<\rangle$ respectively. Thus considering the basis of $\{|0^{+}\rangle, |1^{+}\rangle\}$ in place of the basis $\{|Z\rangle,|Z^{\perp}\rangle\}$, we find that the error $\epsilon$ decreases exponentially with the number of qubits as $e^{-\nu/4}$. Thus $\nu$ is $\Theta\left(\ln \frac{1}{\epsilon}\right)$

	The $U$-complexity of the modified algorithm $\mathcal{N}_{U}(M_{\epsilon})$ is given by $\nu\Theta\left(\mathcal{N}_{U}(P_{1/32})\right)$ which is roughly same as $\mathcal{N}_{U}(H_{\epsilon})$. Note that the multiplication with $\nu$ is due to the fact that simulation by each ancilla qubit takes $\Theta(1)$ applications of $P_{1/32}$ and hence $\Theta\left(\mathcal{N}_{U}(P_{1/32})\right)$ applications of $U$. The ancilla-complexity $\mathcal{N}_{A} (M_{\epsilon})$ is the sum of the $\Theta \left(\ln\frac{1}{\Delta}\right)$ ancilla qubits needed by one register to implement $P_{1/32}$ and $\nu = \Theta\left(\ln\frac{1}{\epsilon}\right)$ ancilla qubits required for majority-voting. This is roughly same as $\mathcal{N}_{A}(P_{\epsilon})$. Thus the operator $M_{\epsilon}$ retains the advantageous features of $H_{\epsilon}$ and $P_{\epsilon}$ but discards their disadvantageous features. 

\section{Fixed-point quantum search based algorithm}
\label{fixedpoint}

	In this section, we show that the majority-voting is not the only way to get an efficient algorithm for the EMP. We present an algorithm based on the combination of the PEA with the fixed-point quantum search algorithms (hereafter referred to as FPQS). Using this, we need only $\mu$ ancilla qubits required by the operator $P_{1/32}$ and we get rid off extra $\nu$ ancilla qubits required for the majority-voting. Thus we get an algorithm which is much easier to implement physically. 

	The FPQS was originally discovered to supplement the standard quantum search algorithm. In the quantum search algorithm, the main goal is to drive the quantum system from an initial source state $|s\rangle$ to an unknown final target state $|t\rangle$ by using selective phase transformations of these two states. The standard quantum search algorithm, though proved to be optimal when $|\langle s|t\rangle| \ll 1$, does not work so well when $|\langle s|t\rangle| \approx 1$ and we need FPQS for a better performance in such cases. The FPQS ensures the monotonic convergence of the source state $|s\rangle$ towards the target state $|t\rangle$. 

	We will be using a procedure which is not exactly same as the FPQS but completely inspired by the FPQS. We write the source state $|s\rangle$ as $V|\sigma\rangle$ where $|\sigma\rangle$ is a standard known quantum state and $V$ can be any unitary operator. We write $V|\sigma\rangle$ as
\begin{equation}
|s\rangle = V|\sigma\rangle = \sqrt{1-\eta^{2}} |t\rangle + \eta |t_{\perp}\rangle, \label{Vsigma}
\end{equation} 
where the non-target state $|t_{\perp}\rangle$ is orthogonal to the target state $|t\rangle$. For any quantum state $|\omega\rangle$ and any angle $\alpha$, the selective phase rotation of $|\omega\rangle$ by an angle of $\alpha$ is given by
\begin{equation}
I_{\omega}^{\alpha} = \mathbbm{1}-(1-e^{\imath \alpha})|\omega\rangle\langle \omega|, \label{SPR}
\end{equation}
where $\mathbbm{1}$ denotes the identity operator.

	The FPQS is based on the following expression. 
\begin{equation}
\left(VI_{\sigma}^{\pi/3}V^{\dagger}I_{t}^{\pi/3}V\right)|\sigma\rangle = \sqrt{1-\eta^{6}}|t\rangle+\eta^{3}|t_{\perp}\rangle. \label{fixedbasic}
\end{equation}
We assume $\eta \ll 1$ so that Eq. (\ref{Vsigma}) can be written as
\begin{equation}
V|\sigma\rangle = \left(1-(0.5)\eta^{2}\right) |t\rangle + \eta |t_{\perp}\rangle. \label{Vsigma1}
\end{equation}
Also, for $\eta \ll 1$, Eq. (\ref{fixedbasic}) can be written as
\begin{equation}
\left(VI_{\sigma}^{\pi/3}V^{\dagger}I_{t}^{\pi/3}V\right)|\sigma\rangle = \left(1-(0.5)\eta^{6}\right)|t\rangle+\eta^{3}|t_{\perp}\rangle. \label{fixedbasic1}
\end{equation}
Note that $|t\rangle$ and $|t_{\perp}\rangle$ are mutually orthogonal and complementary states. Hence the operator $I_{t}^{-\alpha}$ is equivalent (up to an ignorable global phase factor) to the operator $I_{t\perp}^{\alpha}$ which is a selective phase rotation of the $|t_{\perp}\rangle$. Using Eq. (\ref{Vsigma1}) and interchanging the roles of $|t\rangle$ and $|t_{\perp}\rangle$ in Eq. (\ref{fixedbasic1}), we get 
\begin{equation}
\left(VI_{\sigma}^{\pi/3}V^{\dagger}I_{t}^{-\pi/3}V\right)|\sigma\rangle = \left(1-(1.5)\eta^{2}\right) |t\rangle + \sqrt{3}\eta |t_{\perp}\rangle.  \label{fixedbasic2}
\end{equation}
Thus, using $I_{t}^{\pi/3}$, we reduce the amplitude of the $|t_{\perp}\rangle$ exponentially from $\eta$ to $\eta^{3}$, but by using $I_{t}^{-\pi/3}$, we enhance this amplitude linearly from $\eta$ to $\sqrt{3}\eta$.

	We note that Eq. (\ref{P132define}) implies that if the main space is in $|\psi\rangle$ state (the eigenstate to be marked) then the action of $P_{1/32}$ (hereafter denoted as $P$ for simplicity) on the work-space is given by
\begin{equation}
P|\sigma\rangle_{\psi} = \left(1-(0.5)\eta^{2}\right)|t\rangle + \eta|t_{\perp}\rangle,\ \ \eta \leq 2^{-5} \ll 1. \label{P1}
\end{equation}
Here we have written the states $|Z\rangle$ and $|Z^{\perp}\rangle$ as $|t\rangle$ and $|t_{\perp}\rangle$ respectively. The subscript $\psi$ of $|\sigma\rangle$ denotes the state of main space. Similarly if the main space is in any other eigenstate $|\psi_{\perp}\rangle$ then the action of $P$ on the work-space is given by
\begin{equation}
P|\sigma\rangle_{\psi \perp} = \eta|t\rangle + \left(1 - (0.5)\eta^{2}\right)|t_{\perp}\rangle,\ \ \eta \leq 2^{-5} \ll 1. \label{P2}
\end{equation}
We point out that the notation for $\eta$ in Eqs. (\ref{P1}) and (\ref{P2}) does not reflect the fact that its exact value may depend upon the state of the main space. For our purpose, the only relevant fact is that $\eta$ is a positive number with value less than $2^{-5}$. 

	We define an operator $P(1,0)$ as
\begin{equation}
P(1,0) = P(0,0)I_{\sigma}^{\pi/3}P(0,0)^{\dagger}I_{t}^{\pi/3}P(0,0),\ \ P(0,0)  = P. \label{P10define}
\end{equation}
Then putting $V = P(0,0)$ in Eq. (\ref{fixedbasic1}) and using Eqs. (\ref{P1}) and (\ref{P2}), we get
\begin{eqnarray}
P(1,0)|\sigma\rangle_{\psi} & = & \left(1-(0.5)\eta^{6}\right)|t\rangle + \eta^{3}|t_{\perp}\rangle  \nonumber \\
P(1,0)|\sigma\rangle_{\psi \perp} & = & \sqrt{3}\eta|t\rangle + \left(1 - (1.5)\eta^{2}\right)|t_{\perp}\rangle. \label{P10action}
\end{eqnarray} 
The fact $\eta \leq 2^{-5} \ll 1$ has been used in getting above equations. The error term reduces exponentially from $\eta$ to $\eta^{3}$ when the main system is in $|\psi\rangle$ state. But when the main system is in $|\psi_{\perp}\rangle$ state then the error term increases linearly from $\eta$ to $\sqrt{3}\eta$. To reduce this, we define another operator $P(1,1)$ as
\begin{equation}
P(1,1) = P(1,0)I_{\sigma}^{\pi/3}P(1,0)^{\dagger}I_{t}^{-\pi/3}P(1,0).
\end{equation}
Then, using eq. (\ref{fixedbasic2}), we get
\begin{eqnarray}
P(1,1)|\sigma\rangle_{\psi} & = & \left(1-(1.5)\eta^{6}\right)|t\rangle + \sqrt{3}\eta^{3}|t_{\perp}\rangle  \nonumber \\
P(1,1)|\sigma\rangle_{\psi \perp} & = & 3^{3/2}\eta^{3}|t\rangle + \left(1 - (27/2)\eta^{6}\right)|t_{\perp}\rangle. \label{P11action}
\end{eqnarray} 
As desired, the error term reduces from $\eta$ to $\Theta(\eta^{3})$ irrespective of the state of main space.

	We can design a recursive sequence of the operators to further reduce the error term. Let $P(q,q)$ be an operator with the following property
\begin{eqnarray}
P(q,q)|\sigma\rangle_{\psi}  & = & \left(1-(0.5)g_{q}^{2}\eta^{2m_{q}}\right)|t\rangle + g_{q}\eta^{m_{q}}|t_{\perp}\rangle \nonumber \\ 
P(q,q)|\sigma\rangle_{\psi \perp} & = & h_{q}\eta^{m_{q}}|t\rangle + \left(1 - (0.5)h_{q}^{2}\eta^{2m_{q}}\right)|t_{\perp}\rangle. \label{Pqqdefine}
\end{eqnarray}
Here $g_{q}$, $h_{q}$ and $m_{q}$ are some functions of $q$. A comparison with Eqs. (\ref{P1}) and (\ref{P2}) implies that
\begin{equation}
g_{0} = h_{0} = m_{0} = 1. \label{ghminitial}
\end{equation}
We define an operator
\begin{equation}
P(q+1,q) = P(q,q)I_{\sigma}^{\pi/3}P(q,q)^{\dagger}I_{t}^{\pi/3}P(q,q). \label{Pqplusqdefine}
\end{equation}
Then Eq. (\ref{fixedbasic1}) implies that
\begin{eqnarray}
P(q+1,q)|\sigma\rangle_{\psi} & =  & \left(1-(0.5)g_{q}^{6}\eta^{6m_{q}}\right)|t\rangle + g_{q}^{3}\eta^{3m_{q}}|t_{\perp}\rangle \nonumber \\ 
P(q+1,q)|\sigma\rangle_{\psi \perp} & = & \sqrt{3} h_{q}\eta^{m_{q}}|t\rangle + \left(1 - (1.5)h_{q}^{2}\eta^{2m_{q}}\right)|t_{\perp}\rangle. \label{Pqqdefine2}
\end{eqnarray}
Next, we define an operator
\begin{equation}
P(q+1,q+1) = P(q+1,q)I_{\sigma}^{\pi/3}P(q+1,q)^{\dagger}I_{t}^{-\pi/3}P(q+1,q). \label{Pqplusqplusdefine1}
\end{equation}
Then Eq. (\ref{fixedbasic2}) implies that
\begin{eqnarray}
P(q+1,q+1)|\sigma\rangle_{\psi} &=& \left(1-(1.5)g_{q}^{6}\eta^{6m_{q}}\right)|t\rangle + \sqrt{3}g_{q}^{3}\eta^{3m_{q}}|t_{\perp}\rangle \nonumber \\ 
P(q+1,q+1)|\sigma\rangle_{\psi \perp} & = & 3^{3/2} h_{q}^{3}\eta^{3m_{q}}|t\rangle + \left(1 - (27/2)h_{q}^{6}\eta^{6m_{q}}\right)|t_{\perp}\rangle. \label{Pqplusqplusdefine}
\end{eqnarray}
A comparison with Eq. (\ref{Pqqdefine}) gives the following recursive relations for $g_{q}$, $h_{q}$, and $m_{q}$.
\begin{equation}
m_{q+1} = 3m_{q},\ \ g_{q+1} = 3^{1/2} g_{q}^{3},\ \ h_{q+1} = 3^{3/2} h_{q}^{3}.
\end{equation} 
Using Eq. (\ref{ghminitial}), we find the following solutions.
\begin{equation}
m_{q} = 3^{q},\ \ g_{q} = (3^{1/4})^{m_{q}-1},\ \ h_{q} = (3^{3/4})^{m_{q}-1}. 
\end{equation}
Putting this in Eq. (\ref{Pqqdefine}) and using $\eta \leq 2^{-5}$, it is easy to check that
\begin{eqnarray}
P(q,q)|\sigma\rangle_{\psi} & = & |t\rangle + |\epsilon_{q}\rangle, \nonumber \\
P(q,q)|\sigma\rangle_{\psi \perp} & = & |t_{\perp}\rangle + |\epsilon_{q}\rangle,
\end{eqnarray}
where $|\epsilon_{q}\rangle$ denote any state of length less than $\epsilon_{q}$ and
\begin{equation}
\epsilon_{q} = \left(3^{3/4}2^{-5}\right)^{3^{q}} \approx 0.07^{3^{q}}. 
\end{equation}
Thus the error term can be made as small as possible by choosing suitable value of $q$ (the level of recursion). We don't need any extra registers or ancilla qubits to decrease the error as was done in the majority-voting based algorithms. This is a big relief in general situations where spatial resources are constrained. These ancilla qubits were needed to do controlled transformations which are harder to implement physically. Thus the FPQS based algorithm is also much easier to implement physically.

	However, the simplicity of the FPQS-based algorithm comes at the sligtly extra cost of the $U$-complexity. To show this, let $\mathcal{N}_{P}(P(q,q))$ denote the $P$-complexity of the operator $P(q,q)$, i.e. the required number of applications of the operator $P$ or $P^{\dagger}$ by the operator $P(q,q)$. We have $\mathcal{N}_{P}(P(q,q)) = \mathcal{N}_{P}(P(q,q)^{\dagger})$. Eqs. (\ref{Pqplusqdefine}) and (\ref{Pqplusqplusdefine1}) imply that 
\begin{equation}
\mathcal{N}_{P}(P(q+1,q+1)) = 3\mathcal{N}_{P}(P(q+1,q)) = 9\mathcal{N}_{P}(P(q,q)). 
\end{equation}
As $P(0,0)$ is $P$ by definition, $\mathcal{N}_{P}(P(0,0))$ is $1$ and hence the above equation has the following solution.
\begin{equation}
\mathcal{N}_{P}(P(q,q)) = 9^{q} = \Theta\left(\ln^{2}\frac{1}{\epsilon_{q}}\right). 
\end{equation}
Each application of $P$ or $P^{\dagger}$ has an $U$-complexity of $\Theta \left(\ln\frac{1}{\Delta}\right)$. Hence the $U$-complexity of our algorithm is 
\begin{equation}
\Theta\left(\ln\frac{1}{\Delta}\right)\Theta\left(\ln^{2}\frac{1}{\epsilon}\right).
\end{equation}
This is larger than majority-voting based algorithms by a factor of $\Theta\left(\ln\frac{1}{\epsilon}\right)$. This space-time tradeoff can be beneficial if the spatial resources are very constrained or the controlled transformations are physically hard to implement. 

\section{Discussion and Conclusion}
\label{discussion}

	We have presented an algorithm to mark the unknown eigenstates of an operator. The previous algorithms used the concept of majority-voting to enhance the accuracy of the phase-estimation algorithm for this task. We have shown that by using fixed-point quantum search in place of the majority-voting, we can get rid off ancilla qubits and controlled transformations needed for majority-voting.

	We have presented the algorithm for the case of marking only one eigenstate or a degenerate eigenspace with a particular eigenvalue. But the algorithm can be easily extended to the situations when we want to mark multiple eigenstates whose eigenphases lie in a particular interval. This is the situation in the case of quantum principal component analysis where the goal is to mark eigenstates with some of the largest eigenvalues.

	As discussed earlier, marking of the eigenstates is a very important operation in speeding up the general quantum search algorithms using quantum post-processing as shown in ~\cite{postprocessing}. As the general quantum search algorithm finds applications in situations of physical interest ~\cite{spatial,fastersearch,fastergeneral,clause,kato,shenvi,realambainis}, our algorithm also finds applications in those situations. We believe that our algorithm may also find other important applications.

\end{document}